\documentclass[showpacs,aps,prl,longbibliography,showkeys,superscriptaddress,twocolumn]{revtex4-1}

\usepackage[colorlinks, urlcolor=blue,linkcolor=blue, anchorcolor=blue, citecolor=blue]{hyperref}
\usepackage[english]{babel}
\usepackage[utf8]{inputenc}
\usepackage{multirow}
\usepackage{array}
\usepackage{amsthm}
\usepackage{mathtools}
\usepackage{physics}
\usepackage{xcolor}
\usepackage{graphicx}
\usepackage{bm}
\usepackage{soul}
\usepackage{adjustbox}
\usepackage{placeins}
\usepackage[T1]{fontenc}
\usepackage{lipsum}
\usepackage{csquotes}
\usepackage{float}
\usepackage{booktabs}
\usepackage{lineno}
\usepackage{balance}

\graphicspath{{figs/}}

\newcommand{\snn}{\sqrt{s_\mathrm{NN}}}

\newcommand{\vreff}{v_r^\mathrm{eff}}

\newcommand{\Tll}{T_{\ell\bar\ell}}

\newcommand{\Tgz}{T_{\gamma0}}
\newcommand{\Tg}{T_{\gamma}}

\newcommand{\Appendix}{Supplemental Material}

\begin{document}

\title{Electromagnetic Tomography of Radial Flow in the Quark-Gluon Plasma}

\author{Lipei Du}
\email{Corresponding author: ldu2@lbl.gov}
\affiliation{Department of Physics, University of California, Berkeley CA 94270}
\affiliation{Nuclear Science Division, Lawrence Berkeley National Laboratory, Berkeley CA 94270}
\author{Ulrich Heinz}
\affiliation{Department of Physics, The Ohio State University, Columbus, Ohio 43210}

\date{\today}

\begin{abstract}
We present a novel multimessenger approach to extract the effective radial flow of the quark-gluon plasma (QGP) by jointly analyzing thermal photon and dilepton spectra in heavy-ion collisions. A key feature of this method is that it circumvents the need for a directly unmeasurable reference---the photon temperature in the absence of flow---by establishing, within a calibrated model framework, a stable, approximately linear correlation with the dilepton-inferred temperature. This construction defines an experimentally constructible quantity, $v_r^\mathrm{eff}$, which reflects early-time collectivity and exhibits a strong correlation with the spacetime-averaged radial velocity of the QGP. Together with previous results linking dilepton slopes to the initial QGP temperature, our work establishes a consistent framework for electromagnetic tomography of the QGP. Our framework quantifies the experimental precision target, thereby providing a concrete roadmap for future measurements at RHIC and the LHC and opening a new avenue to probe the early-time dynamics of hot QCD matter.
\end{abstract}

\maketitle
{\bf\textit{Introduction.---}}
The quark-gluon plasma (QGP)---a deconfined state of quarks and gluons---is believed to have filled the early universe microseconds after the Big Bang \cite{Braun-Munzinger:2008szb, Shuryak:2014zxa}. Today, relativistic heavy-ion collisions recreate this phase of matter under controlled laboratory conditions, enabling the study of its properties with increasing precision \cite{Muller:2012zq, Bzdak:2019pkr, Du:2024wjm}. A hallmark of QGP evolution is its collective expansion, which is driven by pressure gradients and manifests in strong radial (transverse) flow. Understanding how this flow develops is crucial for revealing the medium’s Equation of State (EoS) and transport properties \cite{Bernhard:2019bmu, JETSCAPE:2020shq, Nijs:2020ors}. The observed non-zero anisotropic flow ($v_n$) of direct photons \cite{PHENIX:2015igl, ALICE:2018dti} has firmly established their sensitivity to the medium’s collective dynamics. This provides strong motivation to extend the inquiry from anisotropic flow to radial flow, which represents a distinct aspect of collectivity: it quantifies the isotropic expansion velocity rather than azimuthal asymmetries. We demonstrate that electromagnetic probes can access this complementary component, defining a realistic precision target for upcoming experimental programs.

We explore a multimessenger strategy in QGP tomography by jointly analyzing thermal photons ($\gamma$) and dileptons ($\ell^+\ell^-$)---electromagnetic probes that escape the medium with minimal final-state interactions \cite{Shuryak:1978ij, Kajantie:1981wg, McLerran1984, Hwa:1985xg, Kajantie:1986dh}. These probes carry comparatively direct, time-dependent information about the evolving medium. While both photons and dileptons are emitted throughout the QGP evolution, they exhibit distinct sensitivities to the medium’s conditions. Photons are blue-shifted by the QGP’s transverse expansion, causing the ``effective temperature'' (inverse slope) extracted from their spectra to appear higher than the actual medium temperature \cite{vanHees:2011vb, Shen:2013vja, Paquet:2015lta, Paquet:2023bdx, Du:2024pbd}. Dilepton invariant-mass spectra, in contrast, are comparatively insensitive to flow effects, offering a cleaner measure of the medium’s intrinsic temperature \cite{Rapp:2014hha, Churchill:2023vpt, Churchill:2023zkk, Massen:2024pnj}. Most thermal photons and dileptons are emitted earlier than hadrons, which undergo strong final-state interactions. This complementarity opens a unique opportunity: by contrasting photon and dilepton spectra, one can disentangle the effects of flow and temperature with tools that are naturally biased toward the earliest and most dynamic stage of expansion.

The central difficulty of this approach is the inaccessibility of the baseline photon spectrum in the absence of flow. We show that this missing reference can be constrained using the dilepton-extracted temperature, exploiting a stable correlation observed across a wide range of beam energies and centralities in simulations calibrated to hadronic observables. This correlation is established for Au+Au collisions spanning nine energies from $\snn=7.7$ to 200 GeV and eight centrality classes from 0--10\% to 70--80\%. Leveraging this correlation, we define an effective radial flow constructed from experimentally measurable electromagnetic signals, whose interpretation is grounded in a calibrated dynamical framework. This enables a novel, data-driven form of electromagnetic tomography of QGP evolution, designed to fully exploit the precision reach of current and upcoming experimental programs.

{\bf\textit{Framework and setup.---}} 
We illustrate our proposed procedure by using realistic model predictions for photons and dileptons obtained from a state-of-the-art multistage hydrodynamic framework \cite{Shen:2020jwv, Du:2023gnv}. This model simulates the spacetime evolution of heavy-ion collisions using a (3+1)-dimensional multistage hydrodynamic framework. The model comprises parametric initial conditions, the MUSIC hydrodynamics code \cite{music1,music2} for QGP evolution, the iS3D sampler \cite{McNelis:2019auj} for particlization, and UrQMD \cite{Bass1998,Bleicher1999} for the hadronic afterburner. The hydrodynamic evolution solves conservation laws with shear stress and baryon diffusion \cite{Denicol2018, Du:2019obx, Shen:2020jwv} but neglecting bulk viscosity \cite{fnbulk}. The fluid is initialized at a fixed \cite{fntau} proper time and evolved until particlization, which is defined on a freezeout surface of constant energy density \cite{Shen:2020jwv, Du:2023gnv}. 

This modeling setup has been validated and its parameters calibrated through systematic comparison with hadronic observables \cite{Du:2022yok, Churchill:2023vpt, Du:2023gnv, Du:2023efk, Du:2024pbd}---including particle yields, transverse momentum spectra, and mean $p_T$ of identified hadrons---both at midrapidity \cite{STAR:2008med, STAR:2017sal, STAR:2019vcp} and forward/backward rapidities \cite{Back:2002wb, PHOBOS:2005zhy, BRAHMS:2009acd, BRAHMS:2009wlg, BRAHMS:2003wwg, BRAHMS:2001llo}, from $\snn = 7.7$ to $200$~GeV. This calibration ensures a realistic and consistent description of the medium across beam energies.

With the spacetime evolution in place, we compute thermal photon and dilepton spectra by integrating their emission rates over the full \cite{fnprehydro} hydrodynamic history \cite{Shen:2013cca, Churchill:2023vpt}. To isolate thermal QGP radiation \cite{Churchill:2023zkk, Du:2024pbd}, only fluid cells with temperatures above the Cleymans freezeout line \cite{Cleymans:2005xv, STAR:2017sal} are considered \cite{fnfrz}. The local rest frame (LRF) emission rates are computed using thermal field theory \cite{Laine:2016hma} via the electromagnetic spectral function $\text{Im}\, \Pi_{\text{em}}(k;\,T, \mu_B)$, derived from the retarded photon self-energy $\Pi_{\text{em}}^{\mu\nu}$ incorporating QCD corrections. In the LRF of each fluid cell with $(T,\,\mu_B)$, we evaluate the photon rate $\omega \, \mathrm{d}^3\Gamma_\gamma/\mathrm{d}^3\mathbf{k}$ \cite{Arnold:2001ms, Gervais:2012wd} and dilepton rate $\mathrm{d}^4\Gamma_{\ell\bar{\ell}}/\mathrm{d}\omega\,\mathrm{d}^3\mathbf{k}$ \cite{Laine:2013vma, Ghisoiu:2014mha, Churchill:2023vpt}, both of which include complete $O(\alpha_s)$ contributions as well as Landau-Pomeranchuk-Migdal effects. The four-momentum $k^\mu{\,=\,}(\omega,\, \mathbf{k})$ satisfies $M{\,=\,}\sqrt{\omega^2 - |\mathbf{k}|^2}$, with invariant mass $M{\,=\,}0$ for photons and $M{\,>\,}0$ for dileptons. 

These rates are then boosted to the laboratory frame using the local four-velocity $u^\mu(x)$ of the fluid. 
We include the dependence on $\mu_B$ in the rates \cite{Gervais:2012wd, Churchill:2023vpt}, but neglect viscous corrections to electromagnetic emission rates in order to maintain a consistent treatment between photons and dileptons, since fully developed viscous corrections for dilepton rates are not yet available. In this sense, the present
framework is intended to define a controlled hydrodynamic baseline for electromagnetic
flow tomography; additional dynamical refinements are naturally incorporated as
systematic improvements upon this baseline, rather than prerequisites for its
construction.

After boosting and integrating over spacetime and phase-space variables \cite{Shen:2013cca, Churchill:2023vpt}, we obtain the dilepton invariant mass spectra $dN_{\ell\bar{\ell}}/(dM\,dy)$ and photon transverse momentum spectra $dN_\gamma/(2\pi p_T\,dp_T\,dy)$. These results have been benchmarked \cite{Churchill:2023vpt, Churchill:2023zkk, Du:2024pbd} against experimental data from the STAR \cite{STAR:2013pwb, STAR2015, STAR:2015zal, STAR:2023wta} and PHENIX \cite{PHENIX:2008uif, PHENIX:2018for, PHENIX:2022rsx} Collaborations, validating the framework’s ability to describe electromagnetic observables. On this basis, a systematic and internally consistent set of photon and dilepton calculations across beam energies and centralities provides the foundation for the following extraction procedure: To extract ``effective temperatures'', we fit the photon $p_T$ spectra at midrapidity within the range $p_T{\,\in\,}[0.8,\,2]$~GeV to an exponential form $\mathrm{d}N_\gamma/(p_T\,\mathrm{d}p_T){\,\propto\,}\exp(-p_T/\Tg)$ \cite{Arnold2001ms, Shen:2013vja, Paquet:2023bdx}. Similarly, the dilepton spectra at midrapidity are fitted to the parametrization $\mathrm{d}N_{\ell\bar{\ell}}/\mathrm{d}M{\,\propto\,}(M\,\Tll)^{3/2} \exp(-M/\Tll)$ for $M{\,\in\,}[1,\,3]$~GeV \cite{Rapp:2014hha, Churchill:2023vpt, Churchill:2023zkk}. Here and throughout, $T_{\ell\bar{\ell}}$, $T_{\gamma}$, and $T_{\gamma 0}$ denote inverse slopes extracted from spectral fits, interpreted as ``effective temperatures'' that characterize the underlying inhomogeneous thermal source rather than the true local fluid temperature. Such slope extractions are already standard practice in dilepton and photon experimental analyses, ensuring direct comparability between our method and existing measurements \cite{HADES:2019auv,STAR:2024bpc,PHENIX:2008uif, PHENIX:2022rsx, ALICE:2015xmh}.

To define the effective radial flow within this framework, we compute photon ``effective temperatures'' with and without \cite{fnvr} transverse expansion, denoted by $\Tg$ and $\Tgz$, respectively. The difference between them reflects the Doppler blue shift induced by radial flow. We then infer the ``effective radial flow velocity'' $v_r^\mathrm{eff}$ using the relativistic blue-shift relation:
\begin{equation}
    \Tg = \Tgz \sqrt{(1 + v_r^\mathrm{eff})/(1 - v_r^\mathrm{eff})}\,.
    \label{blueshift}
\end{equation}
Here, the directly unobservable baseline $\Tgz$ will be inferred from the measurable $T_{\ell\bar{\ell}}$, exploiting the dilepton-photon correlation presented below [Eq.~\eqref{linear}].

\begin{figure}[tp]
    \centering 
    \includegraphics[width=0.7\linewidth]{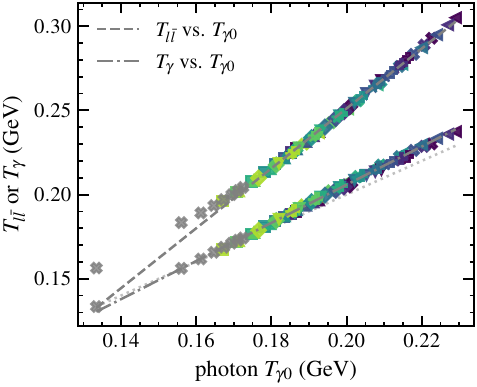}
    \caption{Effective temperatures (see text) at midrapidity extracted from dilepton invariant mass spectra ($\Tll$, upper line) and photon transverse momentum spectra ($\Tg$, lower line), plotted against the photon effective temperature extracted from spectra without transverse flow ($\Tgz$) (see \cite{du_2026_18718103} for a table containing all data points; same for Fig.~\ref{fig2}). Linear fits are shown for each set, excluding the gray points corresponding to $\snn=7.7$~GeV. The dotted line indicates $y = x$.}
    \label{fig1}
\end{figure}

{\bf\textit{Extraction of effective temperatures from electromagnetic spectra.---}}
Figure~\ref{fig1} compares the resulting dilepton and photon inverse slopes ($\Tll$ and $\Tg$) with the one obtained from simulated photons in the absence of transverse expansion ($\Tgz$). We begin by comparing $\Tg$ with $\Tgz$ (lower set of markers in Fig.~\ref{fig1}). As expected, $\Tg$ is systematically higher than $\Tgz$, reflecting the Doppler blue shift of the photon $p_T$-spectra from radial expansion. The difference is small, and it diminishes at lower beam energies, consistent with a less-developed transverse flow in cooler, shorter-lived fireballs; at $\snn=7.7$ GeV they nearly coincide, indicating negligible blue shift.

This systematic shift between $\Tg$ and $\Tgz$ offers a practical route to disentangle flow and temperature effects using Eq.~\eqref{blueshift}. However, because $\Tgz$ is experimentally inaccessible, an alternative non-flow baseline is needed. Motivated by the flow-insensitivity of the dilepton invariant mass spectra \cite{McLerran1984, Rapp:2014hha}, we explore the correlation between $\Tll$ and $\Tgz$. A clear linear trend between them is observed, indicated by the upper line of markers in Fig.~\ref{fig1}, with deviations only at the lowest beam energies and in peripheral collisions (see footnotes~\cite{fntau, fnprehydro}). Interestingly, the integrated $\Tll$ increasingly exceeds $\Tgz$ at higher energies, a consequence of longer lifetimes accentuating their offset \cite{suppl}. Furthermore, $\Tll$ exhibits a steeper slope than $\Tg$ when plotted against $\Tgz$. This reflects the stronger temperature sensitivity of dilepton emission \cite{Churchill:2023vpt,Jackson:2019yao} compared to photons  \cite{Arnold2001ba, Arnold2001ms}: dileptons are preferentially emitted from hotter regions, which skews the extracted temperature upward relative to $\Tgz$ and the blue-shifted $\Tg$.

Crucially, the robust correlation between $\Tll$ and $\Tgz$ allows for the experimental estimation of the otherwise unmeasurable $\Tgz$ using dilepton data. Quantitatively, a global linear fit to the data (excluding $\snn = 7.7$~GeV) gives
\begin{equation}\label{linear}
    \Tll = (1.78\pm0.01)\; \Tgz - (10.5\pm0.2)\times10^{-2}\;\textrm{GeV}\,,
\end{equation}
allowing $\Tgz$ to be inferred from $\Tll$ for use in Eq.~\eqref{blueshift}. In this way, although $v_r^\mathrm{eff}$ is formally defined in terms of the theoretical baseline $\Tgz$, the calibrated $T_{\ell\bar{\ell}}$--$\Tgz$ correlation renders it an experimentally constructible quantity, whose interpretation relies on a calibrated dynamical framework. This forms the core of our data-driven strategy to extract radial flow from electromagnetic probes \cite{fninput}. Dileptons act as baseline thermometers, while photons serve as flow-sensitive blue-shift meters---together enabling the separation of temperature and flow effects that are otherwise entangled in observed spectra. 

We note that at RHIC energies the quantitative difference between $T_\gamma$ and
$T_{\gamma 0}$ is only at the few-percent level, smaller than current experimental
uncertainties. This defines a challenging but well-specified precision target, while existing LHC measurements already point toward a more favorable regime. For example, ALICE measurements of
direct photons in 0--20\% Pb+Pb collisions at $\snn\! =\! 2.76$~TeV report inverse
slopes $T_\gamma \!\sim\! 0.30$~GeV \cite{ALICE:2015xmh}, which---when combined with the calibrated $\Tg$--$\Tgz$ correlation
established here---correspond to a non-flow baseline $T_{\gamma 0} \!\sim\! 0.28$~GeV,
implying a separation of order $20$~MeV. Since the extraction of $v_r^\mathrm{eff}$ depends primarily on the ratio
$T_\gamma/T_{\gamma 0}$, correlated fit and energy-scale uncertainties may
partially cancel, making this separation a conservative benchmark that is
expected to increase at higher collision energies and in more central events
due to hotter and longer-lived fireballs \cite{Marin:2023kqi}.

Looking ahead, ongoing and planned LHC programs, including Runs~3--4 and future ALICE
upgrades \cite{Vertesi:2024piz}, are designed to substantially improve photon and dilepton measurements,
with enhanced statistical precision and reduced backgrounds. In this light, the
present work is not intended to claim immediate experimental extraction, but to
define a new electromagnetic flow observable and to quantify the precision required
to access it. Past experience with electromagnetic probes cautions that the interplay
of temperature, flow, and emission dynamics can reveal effects richer than initially
anticipated, as exemplified by the unexpectedly large photon anisotropic flow
observed at RHIC and the LHC \cite{PHENIX:2015igl, ALICE:2018dti}. By explicitly identifying the relevant scale for radial
flow effects, our framework provides a well-motivated target for future experimental
and theoretical studies.

\begin{figure}[tp]
    \centering 
    \includegraphics[width=0.7\linewidth]{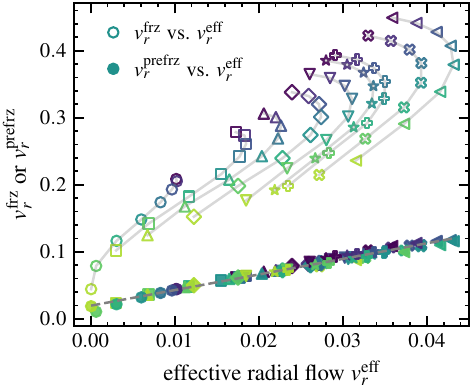}
    \caption{Comparison between the effective radial flow $v_r^\mathrm{eff}$ extracted from electromagnetic spectra and hydrodynamic flow benchmarks: $v_r^\mathrm{prefrz}$ (lower line), the spacetime-averaged radial velocity of fluid cells above the freezeout surface, and $v_r^\mathrm{frz}$ (upper cloud), the averaged flow velocity on the freezeout surface. Points at fixed beam energy but varying centrality are connected by gray lines to guide the eye. A linear fit to $v_r^\mathrm{eff}$ versus $v_r^\mathrm{prefrz}$ is also shown.}
    \label{fig2}
\end{figure}
{\bf\textit{Physical interpretation of the effective photon radial flow $v_r^\mathrm{eff}$.---}}
To give a physical interpretation of the extracted flow observable
$v_r^\mathrm{eff}$ from Eq.~(\ref{blueshift}), we compare it, within the same
calibrated multistage hydrodynamic framework, to other flow quantities obtained
directly from the model. Identifying such correlations is generally beyond experimental reach and only accessible through realistic modeling. Figure~\ref{fig2} presents this comparison, showing $v_r^\mathrm{eff}$ alongside two reference quantities \cite{fnweight}: $v_r^\mathrm{prefrz}$, computed from fluid cells with temperatures above the freezeout line over the entire evolution (closed symbols), and $v_r^\mathrm{frz}$, calculated from fluid cells on the freezeout surface (open symbols).  A strong correlation emerges between $v_r^\mathrm{eff}$ and $v_r^\mathrm{prefrz}$, while no such trend is observed with $v_r^\mathrm{frz}$.

Focusing first on the scatter of $v_r^\mathrm{frz}$ versus $v_r^\mathrm{eff}$, the absence of a universal correlation indicates that $v_r^\mathrm{eff}$ is not simply a reflection of final-state properties near hadronization, as captured by $v_r^\mathrm{frz}$. Interestingly, the centrality dependence of these two flow quantities also differs at fixed beam energy: moving from central to peripheral collisions (top to bottom along each gray line), $v_r^\mathrm{frz}$ (vertical axis) decreases monotonically, consistent with reduced final-state flow in smaller systems. By contrast, $v_r^\mathrm{eff}$ (horizontal axis) increases from central to mid-central collisions, then decreases---revealing a non-monotonic trend.

This behavior is notable: the centrality dependence of $v_r^\mathrm{frz}$ aligns with that of the mean $p_T$ of final hadrons \cite{STAR:2017sal,STAR:2019vcp}, both reflecting the stronger flow built up in hotter, longer-lived central collision fireballs. By contrast, the non-monotonic centrality dependence of $\vreff$, peaking at intermediate centralities, qualitatively mirrors the centrality systematics commonly associated with observables driven by the initial collision geometry, such as hadronic $v_2$ \cite{ALICE:2016ccg, STAR:2017idk}. This behavior is consistent with $\vreff$ being predominantly sensitive to the early-time development of transverse expansion driven by the almond-shaped initial geometry \cite{Heinz:2013th}---a feature captured by electromagnetic emission at early times. That final hadron $v_2$ exhibits a similar trend is not surprising, as it reflects the final anisotropic flow developed throughout the evolution, but rooted in the same initial geometry.

The strong correlation between $v_r^\mathrm{eff}$ and $v_r^\mathrm{prefrz}$ lends this observable clear physical significance. By definition, $v_r^\mathrm{prefrz}$ is the energy-density--weighted radial flow of all fluid cells above the freezeout line, integrated over the entire QGP evolution. That $v_r^\mathrm{eff}$ tracks this quantity so closely supports its interpretation as a spacetime-integrated measure of collective expansion, preceding hadronization. This is expected: electromagnetic probes are continuously emitted throughout the QGP evolution, with a natural bias toward early times when temperatures---and thus emission rates---are highest. In this context, we interpret $v_r^\mathrm{eff}$ as a temperature-weighted, early-time--sensitive radial flow observable. 

We emphasize that $v_r^\mathrm{eff}$ probes the isotropic component of collective
expansion, in contrast to anisotropic flow coefficients that quantify azimuthal
asymmetries. Since non-zero photon anisotropic flow has already been measured
experimentally \cite{PHENIX:2015igl, ALICE:2018dti}, the present work naturally
extends electromagnetic flow studies toward the isotropic radial component,
defining a complementary target for future high-precision measurements.

\begin{figure}[tp]
    \centering 
    \includegraphics[width=0.7\linewidth]{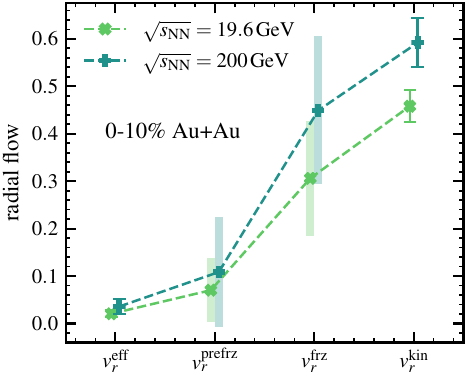}
    \caption{Radial flow quantities at $\snn=19.6$ and $200$~GeV: the effective flow $v_r^\mathrm{eff}$ extracted from electromagnetic spectra, the pre-freezeout average $v_r^\mathrm{prefrz}$ from hydrodynamics, the freezeout surface average $v_r^\mathrm{frz}$, and the kinetic freezeout velocity $v_r^\mathrm{kin}$ obtained from blast-wave fits to hadronic spectra. Error bars indicate fitting uncertainties for $v_r^\mathrm{eff}$ and $v_r^\mathrm{kin}$, while shaded boxes represent the spatial variations among fluid cells for $v_r^\mathrm{prefrz}$ and $v_r^\mathrm{frz}$.}
    \label{fig3}
\end{figure}

{\bf\textit{Time sensitivity of flow quantities and complementarity.---}}
To further illustrate the complementarity between electromagnetic and hadronic flow observables, Figure~\ref{fig3} compares four distinct radial flow quantities at $\snn = 19.6$ and $200$~GeV: the effective radial flow $v_r^\mathrm{eff}$ extracted from electromagnetic probes, the spacetime-averaged pre-freezeout flow $v_r^\mathrm{prefrz}$, the average freezeout surface flow $v_r^\mathrm{frz}$, and the kinetic freezeout velocity $v_r^\mathrm{kin}$ obtained from blast-wave \cite{Schnedermann:1993ws} fits to hadronic spectra \cite{STAR:2017sal}. Over the range shown here all four flow measures increase with beam energy, reflecting the stronger pressure gradients and more explosive expansion of hotter and longer-lived fireballs at higher energies.

A clear hierarchy of magnitudes emerges among the flow quantities at a fixed collision energy, serving as a compact summary of the layered temporal structure of QGP expansion and emission:
\begin{equation}
    v_r^\mathrm{eff} < v_r^\mathrm{prefrz} < v_r^\mathrm{frz} < v_r^\mathrm{kin}.
\end{equation}
Photons and dileptons are emitted predominantly at early times, when the temperature is high but the flow is still modest, biasing their sensitivity toward the onset of collectivity. The hydrodynamic average $v_r^\mathrm{prefrz}$ records the full spacetime development of flow, but remains biased toward early stages due to the temperature dependence of energy weighting. Here, we note that although both are spacetime-averaged over the entire evolution, $v_r^\mathrm{eff}$ is consistently smaller than $v_r^\mathrm{prefrz}$ in magnitude. This difference arises because electromagnetic emission is more heavily weighted toward the hottest regions (i.e. with a higher power of $T$ \cite{Shuryak:1992bt}) than the energy density weightings used in defining $v_r^\mathrm{prefrz}$. In contrast, hadronic observables capture the final-state expansion: $v_r^\mathrm{frz}$ reflects conditions at chemical freezeout, while $v_r^\mathrm{kin}$ additionally accounts for hadronic acceleration between chemical and kinetic freezeout \cite{Petersen:2008dd,Ryu:2017qzn}.

Figure~\ref{fig3} thus encapsulates the continuous development of radial flow, with each quantity providing sensitivity to a different stage of the system’s evolution. Electromagnetic observables thus provide direct access to early-time collective dynamics that are largely inaccessible to hadronic measurements. Although $v_r^\mathrm{eff}$ is numerically small, its temperature-weighted, spacetime-integrated nature biases it toward the hottest stages of the evolution, making it a natural complement to final-state hadronic flow observables. As a result, $v_r^\mathrm{eff}$ offers a promising new handle for constraining early-time dynamics and, potentially, the high-temperature EoS using electromagnetic probes.

{\bf\textit{Summary.---}}
We have introduced a novel multimessenger framework for extracting the effective radial flow of the QGP before its decay into hadrons by combining thermal photon and dilepton measurements in relativistic heavy-ion collisions. This approach exploits the distinct sensitivities of the two probes---dilepton invariant mass spectra serve as a clean thermometer largely unaffected by collective motion, while photon transverse momentum spectra are blue-shifted by flow, acting as a dynamical flowmeter.

A central feature of the method is its circumvention of the unobservable no-flow photon baseline. By establishing a robust, model-calibrated correlation between dilepton-inferred temperatures and the no-flow photon temperature baseline, we reconstruct the spectral blue-shift attributable to transverse expansion. The resulting effective flow observable is significantly smaller in magnitude than final-state hadronic flow, consistent with the early-time dominance of electromagnetic emission, yet it shows a strong correlation with the spacetime-averaged radial velocity of the medium prior to freezeout, giving it a physically meaningful interpretation as a time-integrated measure of QGP expansion dynamics.

Together with earlier findings that relate dilepton temperatures to the initial QGP temperature \cite{Churchill:2023zkk}, this work establishes a unified electromagnetic framework in which dileptons function as thermometers for the early state, while the photon--dilepton comparison yields a flowmeter for early-time collectivity. This enables a new form of electromagnetic tomography that captures the early-time structure of QGP expansion, complementing hadronic observables that are primarily sensitive to later stages.

Applied to current and upcoming datasets at RHIC and especially at the LHC, this method enhances the role of electromagnetic radiation as a precision tool for QGP studies and opens the door to applications in small systems, where the nature of collectivity remains debated. By quantifying the few-percent precision scale at RHIC and its expected enhancement at the LHC, our framework provides a concrete roadmap for future measurements, motivates detector upgrades, and firmly positions electromagnetic tomography as a next-generation tool for precision mapping of early-time QGP dynamics.

\vspace{5mm}
{\it \textbf{Acknowledgements.}---}
L.D. \cite{fnackn} acknowledges useful conversations with Shuzhe Shi. This work was supported in part by the U.S. Department of Energy, Office of Science,
Office of Nuclear Physics, under Grant No.~DE-AC02-05CH11231 (L.D.) and Award
No.~DE-SC0004286 (U.H.).
Computations were made on the computers managed by the Ohio Supercomputer Center \cite{OhioSupercomputerCenter1987}.
\balance
\bibliography{refs}

\begin{thebibliography}{86}
\bibitem [{fnt()}]{fntau}%
  \BibitemOpen
  \href@noop {} {}\bibinfo {note} {Reflecting the longer prehydrodynamic phase
  associated with the extended nuclear interpenetration time, initialization
  time $\tau_0$ increases with decreasing center-of-mass energy. We have
  verified that the correlation presented in Fig.~\ref{fig1} is robust under
  variations in $\tau_0$. Specifically, for $\snn = 7.7$~GeV, decreasing
  $\tau_0$ (i.e. assuming earlier hydrodynamization) shifts the markers in
  Fig.~\ref{fig1} upward and to the right; they remain aligned with the main
  trend, resulting in an improved linear correlation at this collision energy.
  See also \cite{fnprehydro}.}\BibitemShut {Stop}%
\bibitem [{fnp()}]{fnprehydro}%
  \BibitemOpen
  \href@noop {} {}\bibinfo {note} {As noted in \cite{fntau}, the
  prehydrodynamic stage becomes increasingly relevant at lower collision
  energies due to the prolonged nuclear interpenetration process. During this
  stage, continuous energy deposition leads to rapidly rising temperatures,
  peaking once the nuclei pass through each other, accompanied by an evolution
  towards chemical equilibrium. Both effects suppress photon and dilepton
  production during this early phase, limiting its contribution to the final
  spectra. The modest deviation from linearity observed at 7.7\,GeV in
  Fig.~\ref{fig1} may reflect the influence of this presently neglected
  component. Prehydrodynamic emission could affect the lower left regions of
  Figs.~\ref{fig1} and~\ref{fig2} and will therefore be incorporated in future
  work.}\BibitemShut {Stop}%
\bibitem [{fnf()}]{fnfrz}%
  \BibitemOpen
  \href@noop {} {}\bibinfo {note} {In principle, one could use the same
  constant-energy-density surface that defines particlization, which in
  practice lies close to the empirical Cleymans freeze-out line in the
  $(T,\mu_B)$ plane. Since the hydrodynamic model provides direct $(T,\mu_B)$
  for each cell, applying this empirical line as a $\mu_B$-dependent filter
  offers a transparent and practically equivalent way to separate QGP and
  hadronic contributions.}\BibitemShut {Stop}%
\bibitem [{fnv()}]{fnvr}%
  \BibitemOpen
  \href@noop {} {}\bibinfo {note} {To eliminate transverse flow effects we
  modify the four-velocity $u^\mu$ of each fluid cell by setting the spatial
  components $u^x{\,=\,}u^y{\,=\,}0$ and adjusting $u^\tau$ to maintain $u^\mu
  u_\mu = 1$. This modified velocity is then used to compute the photon
  spectra.}\BibitemShut {Stop}%
\bibitem [{fni()}]{fninput}%
  \BibitemOpen
  \href@noop {} {}\bibinfo {note} {We emphasize that using a validated
  theoretical framework to establish such a relationship is a common and
  necessary approach in nuclear physics, analogous to connecting hadronic flow
  coefficients $v_n$ to initial-state geometrical eccentricities, or relating
  final-state multiplicities to the initial entropy normalization.}\BibitemShut
  {Stop}%
\bibitem [{fnw()}]{fnweight}%
  \BibitemOpen
  \href@noop {} {}\bibinfo {note} {Both quantities are defined as weighted
  averages of $u^r/u^t$ using $e \gamma$ as the weight, where $e$ is the local
  energy density and $\gamma = u^t$ is the Lorentz boost factor.}\BibitemShut
  {Stop}%
\end{thebibliography}

\appendix*

\renewcommand{\thefigure}{S\arabic{figure}}
\setcounter{figure}{0} 
\setcounter{equation}{0} 
\renewcommand{\thetable}{\arabic{table}}

\section{\Appendix: Correlation between dilepton and photon temperatures}

In the main text, we leverage the empirical observation that the temperature extracted from the dilepton invariant mass spectrum ($\Tll$) provides a reliable proxy for the unobservable photon temperature in the absence of transverse flow ($\Tgz$), owing to their strong linear correlation shown in Fig.~1. This correlation enables an experimentally feasible determination of the effective radial flow, using the measurable photon temperature $\Tg$ and a model-calibrated estimate of $\Tgz$ derived from $\Tll$.

We find that $\Tll$ increasingly exceeds $\Tgz$ at higher beam energies, which we interpret as a natural consequence of the longer fireball lifetime. Over time, the roughly constant offset $\Delta T = \Tll - \Tgz$ accumulates, leading to a larger deviation in the temperatures derived from time-integrated spectra. This constant offset in time was previously observed in [L. Du, Phys.\,Lett.\,B 861 (2025) 139270], where time-resolved analyses at $\snn = 7.7$ and $14.5$~GeV revealed that $\Tll$ and $\Tgz$, extracted at each time step, evolve in parallel with an approximately constant separation $\Delta T$.

\begin{figure*}[!htbp]
    \centering 
    \includegraphics[width=0.7\textwidth]{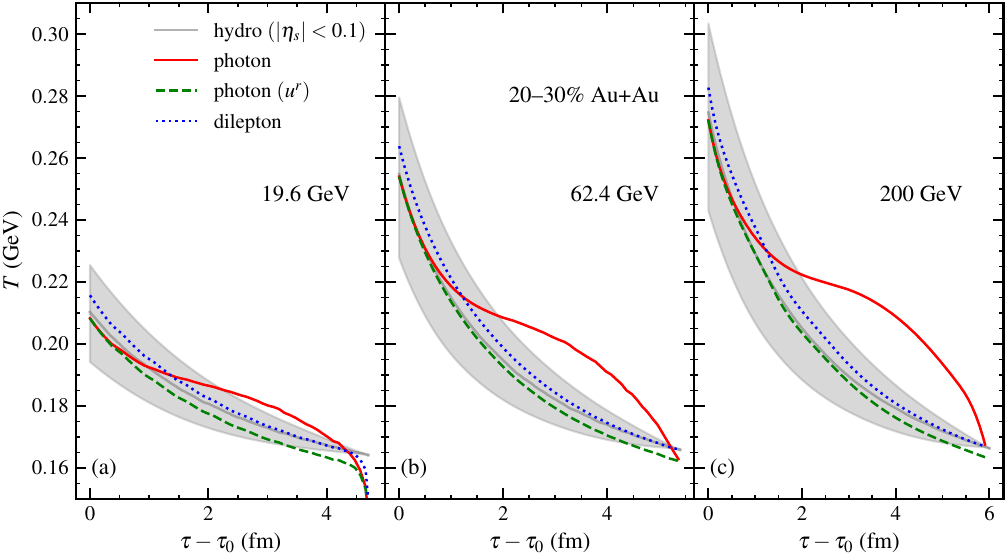}
    \caption{Time evolution of effective temperatures extracted from photon and dilepton spectra at $\snn = 19.6$, $62.4$, and $200$~GeV (panels (a)–(c)). Each panel shows the photon temperature with ($\Tg$, red solid) and without transverse flow ($\Tgz$, green dashed), the dilepton-inferred temperature ($\Tll$, blue dotted), and the hydrodynamic average temperature (gray curve) with its variation (gray band)%
    \footnote{\label{fn:avg}%
    Same as in the main text, an averaged quantity $\langle O \rangle$ is defined as a weighted average over fluid cells with weight $e \gamma$, i.e., $\langle O \rangle = \sum (O \cdot e \gamma)/\sum (e \gamma)$. The gray curve shows this average, while the gray band represents the standard deviation, reflecting the variation among contributing fluid cells. The average radial flow shown in Fig.~\ref{app_fig2} is defined in the same way.}
    in 20–30\% Au+Au collisions at midrapidity. At each energy, $\Tgz$ and $\Tll$ track each other closely throughout the evolution, separated by an approximately constant offset $\Delta T$.}
    \label{app_fig1}
\end{figure*}

\begin{figure*}[!htbp]
    \centering 
    \includegraphics[width=0.7\textwidth]{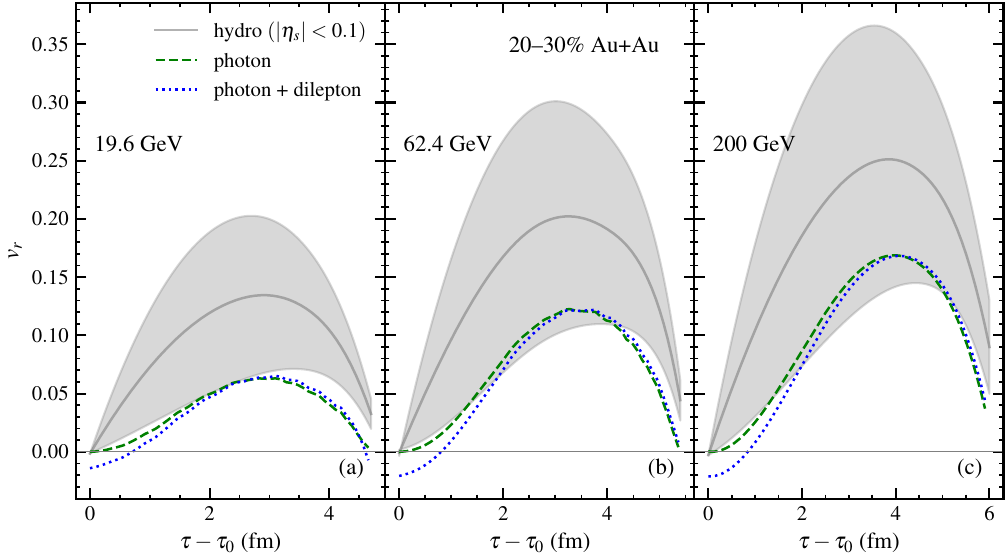}
    \caption{Comparison of effective radial flow extracted using true and proxy baselines for the no-flow photon temperature. For each beam energy shown in Fig.~\ref{app_fig1}, the radial flow $v_r$ is computed from the blue-shift relation using the measured $\Tg$ and either the true $\Tgz$ (green dashed) or the proxy estimate $\Tll - \Delta T$ (blue dotted, denoted as ``photon+dilepton''). The gray curve shows the hydrodynamic average radial flow, and the gray band indicates its standard deviation, reflecting variations among fluid cells weighted as described in footnote~\ref{fn:avg}.}
    \label{app_fig2}
\end{figure*}

To test the generality of this behavior across a broader range of beam energies and centralities, we extend the time-differential analysis to nine energies between $\snn = 7.7$ and $200$~GeV. Figure~\ref{app_fig1} shows the time evolution of $\Tll$ and $\Tgz$ at midrapidity in 20–30\% Au+Au collisions for three representative energies: $\snn = 19.6$, $62.4$, and $200$~GeV. In all cases, $\Tll$ consistently lies above $\Tgz$ throughout the hydrodynamic evolution, with the two curves maintaining a nearly constant offset of a few MeV. Although the overall temperature scale rises with collision energy, the offset $\Delta T$ remains approximately constant during the evolution.

Figure~\ref{app_fig2} evaluates the impact of this approximation on flow extraction. The effective radial flow is computed using the blue-shift relation between $\Tg$ and either (i) the true $\Tgz$ obtained from simulations, or (ii) an adjusted $\Tll$ serving as a proxy for $\Tgz$ via $\Tgz \approx \Tll - \Delta T$. The results demonstrate that the flow extracted using the ``photon+dilepton'' method closely reproduces the reference result obtained with both photon temperatures. The agreement holds across energies and persists throughout the evolution, with small deviations appearing only at early and late times.

Across all nine beam energies and eight centralities, we find that a constant offset $\Delta T = 4.5$~MeV reproduces $\Tgz$ from $\Tll$ for $\snn > 7.7$~GeV, while a slightly smaller value $\Delta T = 4$~MeV suffices at $\snn = 7.7$~GeV. These findings confirm the robustness of the $\Tll$--$\Tgz$ correlation and its near-constant offset in time. While this time-resolved offset is not directly employed in the main analysis, it provides crucial insight into why $\Tll$ increasingly exceeds $\Tgz$ in integrated spectra at higher collision energies.

\end{document}